\documentclass[11pt,reqno]{amsproc}

\usepackage{amsfonts,amsmath,amssymb,amsthm}
\usepackage{amsrefs}
\usepackage{refcount}
\usepackage{xspace}

\mathchardef\ordinarycolon\mathcode`\:
\mathcode`\:=\string"8000
\def\vcentcolon{\mathrel{\mathop\ordinarycolon}} \begingroup
\catcode`\:=\active \lowercase{\endgroup \let :\vcentcolon }

\allowdisplaybreaks[1]

\newcommand{\<}{\langle}
\renewcommand{\>}{\rangle}

\newcommand{\be}{\begin{equation}}
\newcommand{\ee}{\end{equation}}
\def\ba#1\ea{\begin{align}#1\end{align}}
\newcommand{\nn}{\nonumber\\}

\newcommand{\Z}{{\mathbb Z}}
\newcommand{\setM}{\{0,\ldots,M-1\}}
\newcommand{\ZN}{\Z_N}

\newcommand{\Q}{\mathbb{Q}}
\newcommand{\N}{\mathbb{N}}

\renewcommand{\i}{\mathrm{i}}
\newcommand{\ket}[1]{|#1\>}

\newcommand{\E}{\mathop{\mbox{$\mathbf{E}$}}}
\newcommand{\poly}{\mathrm{poly}}
\newcommand{\tr}{\mathop{\mathrm{tr}}\nolimits}
\newcommand{\Prsuc}{\Pr(\text{\rm success})}

\newcommand{\HSP}{\textsc{hsp}\xspace}
\newcommand{\PGM}{\textsc{pgm}\xspace}
\newcommand{\POVM}{\textsc{povm}\xspace}

\newtheorem{theorem}{Theorem}
\newtheorem{lemma}[theorem]{Lemma}

\begin{document}


\title[Quantum algorithm for a generalized hidden shift problem]
      {Quantum algorithm for a \\ generalized hidden shift problem}

\author{Andrew M. Childs}
\address{Andrew M. Childs \newline
 \indent Institute for Quantum Information \newline
 \indent California Institute of Technology \newline
 \indent Pasadena, CA 91125, USA}
\email{amchilds@caltech.edu}

\author{Wim van Dam}
\address{Wim van Dam \newline
 \indent Department of Computer Science \newline
 \indent University of California, Santa Barbara \newline
 \indent Santa Barbara, CA 93106, USA}
\email{vandam@cs.ucsb.edu}


\begin{abstract}
Consider the following generalized hidden shift problem: given a
function $f$ on $\setM \times \ZN$ satisfying $f(b,x)=f(b+1,x+s)$ for
$b=0,1,\ldots,M-2$, find the unknown shift $s\in\ZN$.  For $M=N$, this
problem is an instance of the abelian hidden subgroup problem, which
can be solved efficiently on a quantum computer, whereas for $M=2$, it
is equivalent to the dihedral hidden subgroup problem, for which no
efficient algorithm is known.  For any fixed positive $\epsilon$, we
give an efficient (i.e., $\poly(\log N)$) quantum algorithm for this
problem provided $M\geq N^\epsilon$.  The algorithm is based on the
``pretty good measurement'' and uses H.~Lenstra's (classical)
algorithm for integer programming as a subroutine.
\end{abstract}

\maketitle
\thispagestyle{empty}

\section{Introduction}

Quantum mechanical computers can solve certain problems asymptotically
faster than classical computers, but the extent of this advantage is
not well understood.  The most significant example of quantum
computational speedup, Shor's algorithm for factoring and calculating
discrete logarithms \cite{Sho97}, is essentially based on an efficient
quantum algorithm for the abelian hidden subgroup problem.  This
naturally leads to the question of whether the general {\em
nonabelian} hidden subgroup problem can be solved efficiently on a
quantum computer.  Although efficient algorithms are known for a
number of special cases of this problem
\cites{IMS03,FIMSS03,MRRS04,GSVV04,Gav04,IG04,BCD05b}, the two cases
known to have significant applications, the dihedral group and the
symmetric group, remain unsolved.  In particular, an efficient quantum
algorithm for the hidden subgroup problem (\HSP) over the symmetric
group would lead to an efficient quantum algorithm for graph
isomorphism \cites{BL95,EH99}, and an efficient quantum algorithm for
the dihedral \HSP would lead to efficient quantum algorithms for
certain lattice problems \cite{Reg02}.

Although no polynomial-time algorithm is known for the dihedral \HSP,
Kuperberg recently discovered a subexponential-time quantum algorithm
\cite{Kup03}.  Kuperberg's algorithm uses a superpolynomial amount of
time, space, and queries; Regev subsequently improved the space
requirement to be only polynomial \cite{Reg04}.  These algorithms are
closely related to a connection between the dihedral \HSP and an
average case subset sum problem observed by Regev \cite{Reg02}.

Recently, together with Bacon, we have developed an approach to the
hidden subgroup problem based on the ``{pretty good measurement}''
(\PGM) \cites{BCD05a,BCD05b}.  In this approach, the \PGM is used to
distinguish the members of an ensemble of quantum states corresponding
to the various possible hidden subgroups.  For a variety of groups
that can be written as the semidirect product of an abelian group and
a cyclic group of prime order, we found that this measurement is
closely related to a certain kind of average case algebraic problem.
In particular, the measurement succeeds when the algebraic problem is
likely to have a solution, and can be implemented if the solutions to
that problem can be found.  For the dihedral group, this problem is
simply the average case subset sum problem \cite{BCD05a}; more
generally, we refer to it as the \emph{matrix sum problem}.  In some
cases, the matrix sum problem can be solved, giving an efficient
quantum algorithm for the corresponding hidden subgroup problem
\cite{BCD05b}.  However, since the average case subset sum problem
appears to be difficult, this approach has not yielded an improved
algorithm for the dihedral \HSP.

In this article, we show how the \PGM approach provides an efficient
quantum algorithm for a problem that interpolates between the abelian
and dihedral hidden subgroup problems.  The dihedral \HSP is
equivalent to the \emph{hidden shift problem}, in which the goal is to
determine a hidden shift $s \in \ZN$ given two injective functions
$f_0,f_1$ satisfying $f_0(x)=f_1(x+s)$.  Instead of only two such
functions, we consider $M$ such functions, each one shifted from the
previous by a fixed hidden shift $s$.  If $M=N$, this problem is an
instance of the abelian \HSP on $\ZN\times\ZN$ with the hidden
subgroup $\langle(1,s)\rangle$, which can be solved efficiently using
abelian Fourier sampling.  Even the case $M=N$ is classically
intractable, and the problem only becomes more difficult for smaller
$M$.  In particular, for $M \ll N$, abelian Fourier sampling fails to
determine the hidden shift.  Using the \PGM approach, we give, for any
fixed integer $k \ge 3$, an efficient quantum algorithm that solves
this problem for $M = \lfloor N^{1/k} \rfloor$.  The algorithm works
by implementing a joint measurement on $k$ copies of certain quantum
states that encode the hidden shift.  Because for each $M \geq M'$ the
generalized hidden shift problem on $\{0,\dots,M-1\}\times \ZN$ can be
reduced to the generalized hidden shift problem on
$\{0,\dots,M'-1\}\times \ZN$, for any fixed $\epsilon > 0$, this gives
an efficient quantum algorithm for all $M \geq N^\epsilon$.

By applying the general \PGM approach developed in \cite{BCD05b}, we
find that the matrix sum problem corresponding to the generalized
hidden shift problem is the following: given uniformly random $x \in
\ZN^k$ and $w \in \ZN$, find $b \in \setM^k$ such that $b \cdot x
\bmod N = w$.  As this is a linear Diophantine equation with convex
constraints, it is an instance of integer programming, and can be
solved using Hendrik Lenstra's algorithm for that problem
\cite{Len83}, which is efficient as long as the dimension $k$ is
constant.  Thus our algorithm for the generalized hidden shift problem
reiterates a theme of \cite{BCD05b}: by combining abelian quantum
Fourier transforms with nontrivial classical (or quantum) algorithms,
one can find efficient quantum algorithms for \HSP-like problems via
the implementation of entangled quantum measurements.  This result is
encouraging since entangled measurements are known to be necessary for
some hidden subgroup problems---in particular, for the symmetric group
\cite{MRS05}.

The remainder of this article is organized as follows.  In
Section~\ref{sec:shift}, we describe the generalized hidden shift
problem in detail and explain how it can be viewed as a quantum state
distinguishability problem.  In Section~\ref{sec:pgm}, we review the
pretty good measurement approach to such problems, prove that this
approach solves the generalized hidden shift problem when the number
of states is $k \ge \lceil 1/\epsilon \rceil$, and explain how it can
be implemented by solving an appropriate matrix sum problem.  In
Section~\ref{sec:msp}, we explain how the matrix sum problem can be
solved efficiently (for constant $k$) using Lenstra's algorithm for
integer programming.  Finally, we conclude in
Section~\ref{sec:discussion} with a discussion of the results and some
open questions.

\section{The generalized hidden shift problem}
\label{sec:shift}

It is well known that the dihedral \HSP is equivalent to the {\em
hidden shift problem}, which is defined as follows.  Given two
injective functions $f_0: \ZN \to S$ and $f_1:\ZN \to S$ (where $S$ is
some finite set) satisfying $f_0(x)=f_1(x+s)$ for some unknown $s \in
\ZN$, find $s$.  For a proof of this equivalence, see Theorem 2 of
\cite{EH00} and Proposition 6.1 of \cite{Kup03}.  For certain explicit
functions of interest, such as the Legendre symbol, the hidden shift
problem can be solved efficiently on a quantum computer \cite{DHI03}.
However, for arbitrary black box functions, no efficient algorithm for
the hidden shift problem is known.

A natural generalization of this problem, which we call the {\em
generalized hidden shift problem}, is as follows.  Consider a single
function $f: \setM \times \ZN \to S$ satisfying two conditions: (a) for
fixed $b$, $f(b,x): \ZN \to S$ is injective and (b)
$f(b,x)=f(b+1,x+s)$ for $b=0,1,\ldots,M-2$ for some fixed $s \in \ZN$.
Given such a function, our goal is again to find the hidden shift $s$.
For $M=2$, this problem is simply the usual hidden shift problem (with
$f_b(x)=f(b,x)$ for $b=0,1$), and hence is equivalent to the dihedral
\HSP.  For $M=N$, this problem is an instance of the abelian hidden
subgroup problem (where the hidden subgroup is $\<(1,s)\> \le \ZN
\times \ZN$).  As a step toward understanding the dihedral \HSP, we
would like to investigate the difficulty of the problem for
intermediate values of $M$.  (Note that for intermediate values of
$M$, the generalized hidden shift problem is apparently not an
instance of the \HSP for any group.)

On a quantum computer, this problem can be turned into a state
distinguishability problem in the same manner as the standard approach
to the hidden subgroup problem.  Prepare a uniform superposition over
all values of $b \in \setM$ and $x \in \ZN$ and then compute the value
of $f(b,x)$, giving the state
\be
  \frac{1}{\sqrt{MN}} \sum_{b=0}^{M-1} \sum_{x \in \ZN} |b,x,f(b,x)\>
\,.
\ee
Then measure the third register, giving the state
\be
  |\phi_{x,s}\> := \frac{1}{\sqrt M} \sum_{b=0}^{M-1} |b,x+bs\>
\ee
for some unknown $x \in \ZN$.  Equivalently, the result is the mixed
state described by the density matrix
\be
  \rho_s := \frac{1}{N} \sum_{x \in \ZN} |\phi_{x,s}\>\<\phi_{x,s}|
\,.
\label{eq:state}
\ee

Using a single copy of the state, we can identify $s$ by the standard
period finding algorithm (e.g., as in Shor's algorithm) only when $M$
is very large.  Given the state $\ket{\phi_{x,s}}$, we can try to find
$s$ by applying the Fourier transform over $\ZN \times \ZN$ to the
two registers, which yields the state
\be
  \frac{1}{N\sqrt{M}} \sum_{y,z\in\ZN} \omega^{xz}
    \sum_{b=0}^{M-1} \omega^{b(y+sz)} \, \ket{y,z}
\,,
\ee
where $\omega:=\exp(2\pi\i/N)$.  In the case $M=N$, this state equals
\be
  \frac{1}{\sqrt{N}} \sum_{z\in\ZN} \omega^{xz} \ket{-sz,z}
\,.
\ee
Measuring in the computational basis, we will observe $(-sz,z)$ for a
uniformly random $z\in\Z_N$.  If $z$ is invertible modulo $N$, which
happens with probability $\Omega(1/\log \log N)$, then we can deduce
$s$ immediately from the values $-sz$ and $z$.  However, in general,
for $M\leq N$, the outcome will only be of the form $(-sz,z)$ with
probability $M/N$.  If $M\leq N^\epsilon$ with $\epsilon < 1$, this
probability is exponentially small in $\log N$, and the approach
fails.  A similar argument shows that an analogous approach using a
Fourier transform over $\Z_M \times \ZN$ followed by a computational
basis measurement also fails for $M \ll N$.  (Note that $\poly(\log
N)$ such classical samples information theoretically determine the
answer even for $M=2$ \cite{EH00}, but it is not known how to process
this data efficiently.)

Instead, we will use $k>1$ states and apply the pretty good
measurement.  To see the connection to the matrix sum problem, it is
helpful to write these states in a different basis.  Fourier
transforming the second register over $\ZN$, we find
\be
  \tilde \rho_s^{\otimes k} 
    = \frac{1}{(MN)^k} \sum_{x \in \ZN^k} \sum_{b,c \in \setM^k} 
      \omega^{(b \cdot x-c \cdot x)s} |b,x\>\<c,x|
\label{eq:kstatesfourier}
\,.
\ee
Now introduce the states
\be
  |S^x_w\> := \frac{1}{\sqrt{\eta^x_w}} \sum_{b \in S^x_w} |b\>
\ee
which are uniform superpositions over the solutions of the matrix sum
problem,
\be
  S^x_w := \{b \in \setM^k: b \cdot x = w \bmod N \}
\,.
\ee
Here the number of solutions is $\eta^x_w := |S^x_w|$.  If there are
no solutions (i.e., if $\eta^x_w=0$), then we define $|S^x_w\> := 0$.
In terms of these states, we can rewrite (\ref{eq:kstatesfourier}) as
\be
  \tilde \rho_s^{\otimes k} 
    = \frac{1}{(MN)^k} \sum_{x \in \ZN^k} \sum_{w,v \in \ZN}
      \omega^{(w-v)s} \sqrt{\eta^x_w \eta^x_v}
      |S^x_w,x\>\<S^x_v,x|
\,.
\label{eq:hsstate}
\ee
Given such a state, we would like to determine the value of $s$.

\section{Pretty good measurement approach}
\label{sec:pgm}

In this section, we review the \PGM approach to distinguishing hidden
subgroup states \cites{BCD05a,BCD05b} as applied to the generalized
hidden shift states (\ref{eq:hsstate}).

The \emph{pretty good measurement} (also known as the square root
measurement or least squares measurement) is a measurement that often
does well at distinguishing the members of an ensemble of quantum
states \cite{HW94} (and in fact is sometimes optimal in a certain
sense).  For an ensemble of states $\{\sigma_j\}$ with equal a priori
probabilities, the pretty good measurement is the \POVM with elements
\be
  E_j := \Sigma^{-1/2} \sigma_j \Sigma^{-1/2}
\ee
where
\be
  \Sigma := \sum_j \sigma_j
\ee
and where the inverse is taken over the support of the ensemble.

For the states (\ref{eq:hsstate}), the \PGM normalization matrix is
\be
  \Sigma =  \frac{N}{(MN)^k} \sum_{x \in \ZN^k} \sum_{w \in \ZN}
            \eta^x_w |S^x_w,x\>\<S^x_w,x|
\,,
\ee
giving \POVM elements
\be
  E_j = \frac{1}{N} \sum_{x \in \ZN^k} \sum_{w,v \in \ZN}
        \omega^{(w-v)j} |S^x_w,x\>\<S^x_v,x|
\,.
\label{eq:pgm}
\ee

The probability of successfully identifying the hidden shift $s$ is
independent of $s$, and is given by
\ba
  \Prsuc
    &:= \tr E_s \tilde\rho_s^{\otimes k} \\
    &=  \frac{1}{M^k N^{k+1}} \sum_{x \in \ZN^k} 
        \bigg( \sum_{w \in \ZN} \sqrt{\eta^x_w} \bigg)^2
\label{eq:sucprob}
\,.
\ea
Using this expression, we can show that the success probability is
appreciable when the matrix sum problem is likely to have a solution.
Specifically, we have

\begin{lemma}[Lemma 2 of \cite{BCD05b}]
\label{lem:sucprob}
If $\Pr(\eta^x_w \ge \alpha) \ge \beta$ for uniformly random $x \in
\ZN^k$ and $w \in \ZN$ (i.e., if most instances of the matrix sum
problem have many solutions), then $\alpha \beta^2 N/M^k \le \Prsuc
\le M^k/N$.
\end{lemma}

\begin{proof}
For the upper bound, we have
\be
  \Prsuc \le \frac{M}{N^{k+1}} \sum_{x \in \ZN^k}
              \bigg( \sum_{w \in \ZN} \eta^x_w \bigg)^2
         =   \frac{M^k}{N}
\ee
since the $\eta$'s are integers and $\sum_{w \in \ZN} \eta^x_w = M^k$
for any $x$.  For the lower bound, we have
\be
  \Prsuc \ge \frac{N}{M^k} \bigg( \frac{1}{N^{k+1}} \sum_{x \in
             \ZN^k} \sum_{w \in \ZN} \sqrt{\eta^x_w} \bigg)^2
\ee
by Cauchy's inequality applied to (\ref{eq:sucprob}). Now
\be
  \frac{1}{N^{k+1}} \sum_{x \in \ZN^k} \sum_{w \in \ZN} \sqrt{\eta^x_w}
  \ge \sqrt\alpha \Pr(\eta^x_w \ge \alpha) \,,
\ee
so by the hypothesis,
$\Prsuc \ge \alpha \beta^2 N/M^k$
as claimed.
\end{proof}

For uniformly random $x \in \ZN^k$ and $w \in \ZN$, the expected
number of matrix sum solutions is
\be
  \mu := \E_{x,w}[\eta^x_w] = \frac{M^k}{N}
\,,
\ee
where we have again used the fact that $\sum_w \eta^x_w = M^k$ for any
$x$.  Thus, we expect the matrix sum problem to typically have no
solutions for $k \ll \log N / \log M$, many solutions for $k \gg \log
N / \log M$, and a constant number of solutions for $k \approx \log N
/ \log M$.  This intuition can be formalized as follows:

\begin{lemma} \label{lem:numsolns}
For the generalized hidden shift problem with $M=\lfloor N^{1/k}
\rfloor$ with $k \ge 3$ and $N$ sufficiently large, $\Pr(1 \le
\eta^x_w \le 4)$ is lower bounded by a constant.
\end{lemma}

A proof is given in the appendix.

Together, Lemmas~\ref{lem:sucprob} and \ref{lem:numsolns} show that
the $\PGM$ has at least a constant probability of successfully
identifying the hidden shift.  In fact, it turns out that the \PGM is
the \POVM that maximizes the probability of successfully determining
$s$ given the states (\ref{eq:hsstate}).  For more details, we refer
the reader to Section~4 of \cite{BCD05a} and Section~4.4 of
\cite{BCD05b}.

Finally, to give an efficient algorithm based on the \PGM, we must
show how the measurement can be implemented efficiently on a quantum
computer.  Such an implementation can be achieved using Neumark's
theorem \cite{Neu43}, which states that any \POVM can be realized by a
unitary transformation $U$ on the system together with an ancilla
followed by a measurement in the computational basis.  For a \POVM
consisting of $N$ rank one operators $F_j=|f_j\>\<f_j|$ in a
$D$-dimensional Hilbert space, $U$ has the block form
\be
  U = \begin{pmatrix}V & X \\ Y & Z\end{pmatrix}
\ee
where the rows of the $N \times D$ matrix $V$ are the $D$-vectors
$\<f_j|$, i.e., $V = \sum_{j=1}^N |j\>\<f_j|$.  (In particular, if all
the vectors $|f_j\>$ have unit length, as will be the case below,
$X=0$.)

Recall from (\ref{eq:pgm}) that the \POVM operators for the \PGM on
hidden subgroup states can be written
\be
  E_j = \sum_{x \in \ZN^k} E^x_j \otimes |x\>\<x|
\ee
where
\be
  E^x_j := |e^x_j\>\<e^x_j|
\ee
with
\be
  |e^x_j\> := \frac{1}{\sqrt{N^{k}}} \sum_{w \in \ZN} \omega^{wj} |S^x_w\>
\,.
\ee
In other words, each $E_j$ is block diagonal, with blocks labeled by
$x \in \ZN^k$, and where each block is rank one.  Thus, the
measurement can be implemented in a straightforward way by first
measuring the block label $x$ and then performing the \POVM
$\{E^x_j\}_{j \in \ZN}$ conditional on the first measurement result.

To implement the \POVM $\{E^x_j\}_{j \in \ZN}$ using Neumark's
theorem, we would like to implement the unitary transformation $U^x$
with the upper left submatrix
\be
  V^x = \frac{1}{\sqrt N} \sum_{j,w \in \ZN} 
        \omega^{-wj} \, |j\>\<S^x_w|
\,.
\ee
It is convenient to perform a Fourier transform (over $\ZN$) on the
left (i.e., on the index $j$), giving a unitary operator $\tilde U^x$
with upper left submatrix
\ba
  \tilde V^x &= \frac{1}{N} \sum_{j,w,v \in \ZN} 
                \omega^{(w-v)j} \, |w\>\<S^x_v| \\
             &= \sum_{w \in \ZN} |w\>\<S^x_w|
\,.
\ea
Therefore, the \PGM can be implemented efficiently if we can
efficiently perform a unitary transformation satisfying
\be
  |w,x\> \mapsto |S^x_w,x\>
\label{eq:qsample}
\ee
for all matrix sum instances $(x,w)$ with $\eta^x_w > 0$.  We refer to
(\ref{eq:qsample}) as \emph{quantum sampling} of solutions to the
matrix sum problem.  If we can efficiently quantum sample from matrix
sum solutions, then by running the circuit in reverse, we can
efficiently implement $\tilde U^x$, and hence the desired measurement.

By applying these unitary transformations directly to the state
(\ref{eq:hsstate}), we can obtain a description of the algorithm
without reference to generalized measurement.  Performing the inverse
of the quantum sampling transformation (\ref{eq:qsample}) followed by
a Fourier transform, we obtain the state
\be
  \rho' := \frac{1}{N^k} \sum_{x \in \ZN^k}
           |\rho'_x,x\>\<\rho'_x,x|
\ee
where
\be
  |\rho'_x\>
  := \frac{1}{\sqrt{M^k N}} \sum_{j,w \in \ZN}
     \omega^{w(s-j)} \sqrt{\eta^x_w} |j\>
\,.
\label{eq:finalstate}
\ee
Roughly speaking, if the distribution of $\eta^x_w$ is close to
uniform, then the sum over $w$ in (\ref{eq:finalstate}) is close to
zero unless $j=s$, so that a measurement of the first register is
likely to yield the hidden shift $s$.

In general, it may be difficult to implement (\ref{eq:qsample})
exactly.  Instead, we may only be able to perform an approximate
quantum sampling transformation satisfying
\be
  |w,x\> \mapsto \begin{cases}
    |S^x_w,x\>   & (x,w) \in Z_{\rm good} \\
    |\mu^x_w,x\> & (x,w) \in Z_{\rm bad}
  \end{cases}
\label{eq:aqsample}
\ee
for some states $|\mu^x_w\>$, where $Z_{\rm good},Z_{\rm bad}$ form a
partition of the matrix sum instances $(x,w)$ for which $\eta^x_w>0$.
The good matrix sum problem instances $(x,w) \in Z_{\rm good}$ are
those for which the quantum sampling can be done correctly.  Assuming
the bad matrix sum instances $(x,w) \in Z_{\rm bad}$ can be
recognized, we can ensure that $\<S^x_w|\mu^x_{w'}\>=0$ for all $x \in
\ZN^k, w,w' \in \ZN$.  Applying the approximate quantum sampling
transformation followed by the Fourier transform gives the state
\be
  \rho'_{\rm apx}
  = \frac{1}{N^k} \sum_{x \in \ZN^k}
    |\rho'_{x,{\rm apx}},x\>\<\rho'_{x,{\rm apx}},x|
\ee
where
\ba
  |\rho'_{x,{\rm apx}}\> := \frac{1}{\sqrt{M^k N}} \sum_{j \in \ZN}
    &\Bigg(
     \sum_{(x,w) \in Z_{\rm good}}
     \omega^{w(s-j)}\sqrt{\eta^x_w} |j\> \nn
   &+ \sum_{(x,w) \in Z_{\rm bad}}
     \omega^{w(s-j)}\sqrt{\eta^x_w} |\nu^x_j\>
    \Bigg)
\ea
for some states $|\nu^x_j\>$ with $\<j|\nu^x_{j'}\>=0$ for all $x \in
\ZN^k, j,j' \in \ZN$ (since $\<S^x_w|\mu^x_{w'}\>=0$ for all $x \in
\ZN^k, w,w' \in \ZN$ and (\ref{eq:aqsample}) is unitary).  The
fidelity between the ideal final state $\rho'$ and the actual final
state $\rho'_{\rm apx}$ resulting from approximate quantum sampling is
thus
\be
  \frac{1}{(MN)^k} \sum_{(x,w) \in Z_{\rm good}} \eta^x_w
\,.
\ee
Now $\eta^x_w > 1$ for all $(x,w) \in Z_{\rm good}$, so if $|Z_{\rm
good}|$ is sufficiently large, the actual final state is close to the
ideal final state, and hence a measurement of the first register
yields the hidden shift $s$ with reasonable probability.  As we will
show in the next section, the instances with $1 \le \eta^x_w \le 4$
can be quantum sampled efficiently (i.e., these instances are good).
Then, letting $M=\lfloor N^{1/k} \rfloor$, Lemma 2 shows that $|Z_{\rm
good}|/N^{k+1}$ is lower bounded by a constant, and thus the fidelity
between $\rho'$ and $\rho'_{\rm apx}$ is lower bounded by a constant,
so that the probability of successfully determining $s$ is lower
bounded by a constant.

\section{Solution of the matrix sum problem}
\label{sec:msp}

Recall that the matrix sum problem for the generalized hidden shift
problem is the following: given $x \in \ZN^k$ and $w \in \ZN$ chosen
uniformly at random, find $b \in \setM^k$ such that $b \cdot x =w \bmod N$.  
This is a linear equation over $\ZN$ in $k$ variables, where the
solutions are required to come from $\setM$.  Such solutions can be
found using integer programming, which has an efficient algorithm if
$M$ is sufficiently large.  We assume $M=\lfloor N^{1/c} \rfloor$ for
some positive integer $c \ge 3$.  Since we can always decrease $M$ by
only considering a subset of the inputs to the first argument of the
hiding function $f$, this will not constitute a loss of generality.
According to Lemma~\ref{lem:numsolns}, we take $k=c$ so that there are
between $1$ and $4$ solutions with probability at least some constant.

To see the connection to integer programming, we note that the
solutions form an integer lattice.  We begin by considering the
equation $b\cdot x=w\bmod{N}$ as a $(k+1)$-variable linear equation
over all the integers $\Z$.  Define an extension of $x$ by $\bar
x:=(x_1,\dots,x_k,N)$ and consider the solutions $\bar b\in\Z^{k+1}$
of the equation $\bar b\cdot \bar x = w$.  For any $b\in\Z^k$ that
solves the equation $b \cdot x = w\bmod{N}$, there is a unique
$\lambda \in \Z$ such that $\bar b=(b,\lambda)$ is a solution to $\bar
b \cdot \bar x=w$; and conversely, for any $\bar b \in \Z^{k+1}$ that
solves $\bar b \cdot \bar x=w$, there is a unique $b \in \Z^k$
(namely, the first $k$ components of $\bar b$) that solves $b \cdot x
= w \bmod N$.  Hence there is a bijection between the solutions $\bar
b \in \Z^{k+1}$ to the equation $\bar b\cdot \bar x=w$ and the
solutions $b \in \Z^k$ to the equation $b\cdot x = w\bmod{N}$.

By Lemma~\ref{lem:gcd} in the appendix, we see that the linear
Diophantine equation $\bar b\cdot \bar x=w$ will have no solutions if
$\gcd(x_1,\ldots,x_k,N)$ does not divide $w$.  If $\bar b\cdot \bar
x=w$ does have a solution, then the solutions $\bar b$ comprise a
shifted $k$-dimensional lattice $\bar b^{(0)}+L$ with some particular
solution $\bar b^{(0)}$ satisfying $\bar b^{(0)} \cdot \bar x=w$ and
the elements of $L\subset \Z^{k+1}$ the solutions of the equation
$\bar b \cdot \bar x = 0$.  By omitting the last coordinate of these
solutions, we obtain all solutions $b\in\Z^k$, which comprise a
shifted $k$-dimensional lattice in $\Z^k$: 
\be \label{eq:soln-lattice} 
  b =  b^{(0)} +\sum_{j=1}^{k} \beta_j b^{(j)}
\ee 
for all $\beta_1,\dots,\beta_k\in\Z$.  Due to the aforementioned
bijection, each solution $b \in \Z^k$ has a unique set of coordinates
$\beta_1,\dots,\beta_k \in \Z$.  The vectors $b^{(0)},\dots,b^{(k)}
\in \Z^k$ can be found efficiently by applying the extended Euclidean
algorithm to the equation $\bar b\cdot \bar x=w$ (see for example
Algorithm 1.3.6 in \cite{Coh93}).

To solve the matrix sum problem, we would like to find the solutions
$b$ that lie in $\setM^k$, which is the set of integer points in the
convex region described by the inequalities
\be
  0 \le b_i \le M-1 \,, \quad i=1,\ldots,k
\label{eq:ineqs}
\,.
\ee
The problem of finding such points (or more precisely, deciding
whether such a point exists) is simply an instance of integer
programming in $k$ dimensions, which can be solved efficiently if $k$
is a constant.  In general, the integer programming problem is the
following.  Given a rational matrix $A \in \Q^{m \times k}$ and a
rational vector $\gamma \in \Q^m$, does there exist an integral vector
$\beta \in \Z^k$ such that $A \beta \le \gamma$?  Although this
general problem is NP-complete \cites{Kar72,BT76}, if the dimension
$k$ is held constant, then the problem can be solved in time
polynomial in the input size \cite{Len83} using an algorithm based on
lattice basis reduction \cite{LLL82}.

By rewriting the convex constraints (\ref{eq:ineqs}) in terms of the
lattice of solutions (\ref{eq:soln-lattice}), we see that solutions of
the matrix sum problem correspond precisely to vectors $\beta \in
\Z^k$ satisfying the constraints
\begin{alignat}{2}
   \sum_{j=1}^k \beta_j b^{(j)}_{i} &\le (M-1) - b^{(0)}_{i}
  \,, &\quad i&=1,\ldots,k \\
  -\sum_{j=1}^k \beta_j b^{(j)}_{i} &\le b^{(0)}_{i}
  \,, &      i&=1,\ldots,k
\,.
\end{alignat}
But this is precisely an instance of integer programming in $k$
dimensions with $m=2k$ constraints, with
\ba
  A_{ij}   & = \begin{cases}
                 b^{(j)}_{i} & i=1,\ldots,k \\
                -b^{(j)}_{i-k} & i=k+1,\ldots,2k
              \end{cases} \\
  \gamma_i &= \begin{cases}
                 (M-1) - b^{(0)}_{i}    & i=1,\ldots,k \\
                         b^{(0)}_{i-k} & i=k+1,\ldots,2k \,.
              \end{cases}
\ea
Therefore, it can be solved efficiently whenever $k$ is a constant.
Note that integer programming as described above is a decision
problem, whereas we would like to find the actual solutions.  However,
this is easily accomplished using bisection, recursively dividing the
set $\setM^k$ into halves, to find all of the solutions efficiently
(for the cases in which there are few solutions---in particular, for
those for which there are between $1$ and $4$ solutions).

Overall, we see that we can efficiently solve the matrix sum problem
(in a regime where the pretty good measurement solves the generalized
hidden shift problem with constant probability) whenever $M\geq
N^\epsilon$ for some fixed $\epsilon>0$.  For the cases in which the
number of solutions is small, they can be explicitly enumerated, and
hence we can efficiently perform the approximate quantum sampling
transformation (\ref{eq:aqsample}) (see for example footnote 2 of
\cite{BCD05b}).  Therefore, we find the following result.

\begin{theorem}\label{thm:main}
The generalized hidden shift problem with $M\geq N^\epsilon$ for any fixed
$\epsilon>0$ can be solved in time $\poly(\log N)$ on a quantum
computer.
\end{theorem}

\begin{proof}
Given $\epsilon$, we will use $k=\max\{\lceil 1/\epsilon \rceil,3\}$
copies of the unknown quantum state (\ref{eq:state}).  Because the
generalized hidden shift problem on the full domain $\setM\times\ZN$
can be solved by solving the same problem on a reduced domain
$\{0,\dots,M'-1\}\times\ZN$ with $M'\leq M$, it is sufficient to prove
the theorem for a specific $M\leq N^\epsilon$.  We will do this for
$M=\lfloor N^{1/k}\rfloor$. 

By Lemma~\ref{lem:numsolns}, there is a constant probability of having
between $1$ and $4$ solutions to the random matrix sum equation
$b\cdot x=w\bmod{N}$, and hence the success probability of the \PGM is
also a constant (by Lemma~\ref{lem:sucprob}).  The efficient
approximate implementation of this measurement is described in the
second half of Section~\ref{sec:pgm} in combination with the results
on integer programming that are described in the first part of
Section~\ref{sec:msp}.
\end{proof}

In fact, the algorithm remains efficient even if $\epsilon$ decreases
(very) slowly with $N$.  Lenstra's algorithm for integer programming
in dimension $k$ takes time $2^{O(k^3)}$ \cite{Len83}, so the
generalized hidden shift problem can be solved efficiently for
$M=N^{O(1/(\log \log N)^{1/3})}$.  Indeed, a subsequent improvement by
Kannan solves $k$-dimensional integer programming in time $2^{O(k \log
k)}$ \cite{Kan87}, which can be used to decrease $M$ slightly
further.

\section{Discussion}
\label{sec:discussion}

We have applied the \PGM approach to the generalized hidden shift
problem, which interpolates from the dihedral \HSP to the abelian
$\HSP$ as $M$ varies from $2$ to $N$.  We found an efficient quantum
algorithm for this problem for any $M\geq N^\epsilon$ with $\epsilon$
fixed (or decreasing very slowly with $N$).  The algorithm works by
solving the matrix sum problem using Lenstra's algorithm for integer
programming in constant dimensions, thereby illustrating (as in
\cite{BCD05b}) that nontrivial classical algorithms can be useful for
implementing entangled measurements to distinguish states obtained by
Fourier sampling.

Our original motivation for studying this problem was the observation
that a solution to the generalized hidden shift problem for
sufficiently small $M$ could lead to new algorithms for the unique
shortest vector in a lattice problem, just as Regev showed for the
case $M=2$ \cite{Reg02}.  Unfortunately, $M\geq N^\epsilon$ does not
appear to be sufficiently small to yield interesting lattice
algorithms.  Nevertheless, attempting to solve the generalized hidden
shift problem for yet smaller $M$ may be a promising path toward
improved quantum algorithms for lattice problems.  Indeed, for the
case $M=2$, Kuperberg's subexponential-time algorithm outperforms the
algorithm given in this paper, so it seems likely that an improved
algorithm could be found for values of $M$ intermediate between $2$
and $N^\epsilon$.

Another problem suggested by this work is the following generalization
of graph isomorphism.  Suppose that we are given a list of $n$-vertex
graphs $G_0,\ldots,G_{M-1}$, and are promised that either no two
graphs are isomorphic, or $G_b = \pi(G_{b+1})$ for some fixed
permutation $\pi \in S_n$ for $b=0,1,\ldots,M-2$.  It would be
interesting to show that this problem can be solved efficiently even
for very large $M$ (where the graphs can be specified by a black box
in the case where $M$ is superpolynomial in $n$). 

\section*{Acknowledgments}

We thank Oded Regev for for discussions about the relationship between
lattice problems and the generalized hidden shift problem.
AMC received support from the National Science Foundation under Grant
No.\ EIA-0086038.
WvD's work was supported in part by the  Advanced Research and Development 
Activity (ARDA).

\appendix
\section*{Appendix: Number of solutions of the matrix sum problem}

In this section, we prove Lemma~\ref{lem:numsolns}.  Before giving the
proof, we need the following fact:

\begin{lemma}
\label{lem:gcd}
For any fixed $b$, the number of solutions $x \in \ZN^k$ to the
equation $b \cdot x = 0 \bmod N$ is $N^{k-1} \gcd(b_1,\ldots,b_k,N)$.
\end{lemma}
\begin{proof}
First, consider the case where $N=p^r$ is a prime power.  Then
$\gcd(b_1,\ldots,b_k,p^r)=p^s$ for some $s \in \{0,1,\ldots,r\}$.  In
particular, there must be an index $i$ such that $\gcd(b_i,p^r)=p^s$,
and hence $b_i=cp^s$ for some $c \in \{1,\ldots,p-1\}$.  Without loss
of generality, assume $i=1$.  Now we can rewrite the equation $b \cdot
x = 0$ as $cp^s x_1 + \sum_{j=2}^k b_j x_j = 0 \bmod{p^r}$, or
equivalently, since $p^s$ is a common divisor of all $b_j$, $c x_1 =
-\sum_{j=2}^k b_j' x_j \bmod{p^{r-s}}$ where $b_j' = b_j / p^s$.
Because $c \in \Z_{p^r}^\times$, for any fixed $(x_2,\ldots,x_k) \in
\Z_{p^r}^{k-1}$, there are $p^s$ solutions $x_1 = (\sum_{j=2}^k
b_j'x_j)/c + \lambda p^{r-s} \bmod{p^r}$ (one solution for for each
$\lambda\in\{0,\dots,p^s-1\}$).  Hence the total number of solutions
$(x_1,\ldots,x_k)$ is $N^{k-1}p^s$, proving the lemma for the case
$N=p^r$.

Now if $N$ is not a prime power, let $N=p_1^{r_1} \cdots p_t^{r_t}$ be
the factorization of $N$ into powers of distinct primes, and let
$\tau:\ZN \to \Z_{p_1^{r_1}} \times\cdots\times \Z_{p_t^{r_t}}$ be the
ring isomorphism provided by the Chinese remainder theorem: for $x \in
\ZN$, $\tau(x) = (x \bmod p_1^{r_1},\ldots,x \bmod p_t^{r_t})$.  Since
$\tau$ is a ring isomorphism, $b \cdot x = 0 \bmod N$ if and only if
$b \cdot \tau(x)_i = 0 \bmod p_i^{r_i}$ for all $i=1,\ldots,t$.  By
the special case of the lemma for $N$ a prime power, the number of
solutions to the $i$th equation is $p_i^{r_i(k-1)}
\gcd(b_1,\ldots,b_k,p_i^{r_i})$; hence the total number of solutions
is $\prod_{i=1}^t p_i^{r_i(k-1)} \gcd(b_1,\ldots,b_k,p_i^{r_i}) =
N^{k-1} \gcd(b_1,\ldots,b_k,N)$ as claimed.
\end{proof}

Now we are ready to give the proof of Lemma~\ref{lem:numsolns}:
\setcounterref{theorem}{lem:numsolns}
\addtocounter{theorem}{-1}
\begin{lemma} 
For the generalized hidden shift problem with $M=\lfloor N^{1/k}
\rfloor$ with $k \ge 3$ and $N$ sufficiently large, $\Pr(1 \le
\eta^x_w \le 4)$ is lower bounded by a constant.
\end{lemma}
\begin{proof}
The main idea of the proof is the same as in the proof of Lemma~5 of
\cite{BCD05b}:  we show that the variance of $\eta^x_w$ is small, so
that the number of solutions of the matrix sum problem is typically
close to its mean.  Because we have $M = \lfloor N^{1/k}\rfloor$, the
mean value of $\eta^x_w$ is $\mu := \E_{x,w}[\eta_w^x] = M^k/N =
1+O(1/N)$ as $N$ grows. 

The variance of the number of solutions $b \in \setM^k$ of the
equation $b \cdot x = w \bmod N$ for uniformly random $x \in \ZN^k, w
\in \ZN$ is $\sigma^2 := \E_{x,w}[(\eta^x_w)^2]-\mu^2$, and
\ba
  \E_{x,w}[(\eta^x_w)^2]
    &= \frac{1}{N^{k+1}} \sum_{x\in\ZN^k,w\in\ZN} (\eta^x_w)^2 \\
    &= \frac{1}{N^{k+1}} \sum_{x\in\ZN^k,w\in\ZN}
       \left(\sum_{b} \delta_{b \cdot x,w}\right) 
       \left(\sum_{c} \delta_{c \cdot x,w}\right) \\
    &= \frac{1}{N^{k+1}} \sum_{x,w}
       \left( \sum_{b} \delta_{b \cdot x,w}
             +\sum_{b \ne c} \delta_{b \cdot x,w} \,
                             \delta_{b \cdot x,c \cdot x} \right)
\intertext{(with the $b,c$ summations over $\setM^k$).  The first
(diagonal) term is just the mean.  To handle the second (off-diagonal)
term, we can write}
  \E_{x,w}[(\eta^x_w)^2]
    &=   \mu + \frac{1}{N^{k+1}} \sum_{b \ne c} \sum_{x\in\ZN^k}
         \delta_{b \cdot x,c \cdot x} \sum_{w\in\ZN} \delta_{b \cdot x,w} \\
    &=   \mu + \frac{1}{N^{k+1}} \sum_{b \ne c} \sum_{x\in\ZN^k}
         \delta_{b \cdot x,c \cdot x} \\
    &=   \mu + \frac{1}{N^2} \sum_{b \ne c}
         \gcd(b_1-c_1,\ldots,b_k-c_k,N) \label{eq:varlastexact} \\
    &\le \mu + \frac{1}{N^2} \sum_{b \ne c}
         \gcd(b_1-c_1,\ldots,b_k-c_k)
\intertext{where the next to last step follows from
Lemma~\ref{lem:gcd}.  Now for any $q \in \setM$, for a fixed value of
$c_i$, there are $1 + \lfloor (M-c_i-1)/q \rfloor + \lfloor c_i/q
\rfloor \le 1 + M/q$ choices of $b_i \in \setM$ that are divisible by
$q$, and hence the number of $b,c$ such that
$\gcd(b_1-c_1,\ldots,b_k-c_k)=q$ is upper bounded by $M^k[(M+q)/q]^k$.
Therefore, for fixed $k\geq 3$ and $M^k/N =1 +O(1/N)$, we have}
  \E_{x,w}[(\eta^x_w)^2]
    &\le \mu + \frac{M^k}{N^2} 
               \sum_{q=1}^{M-1} q \bigg(\frac{M+q}{q}\bigg)^k \\
    &=   \mu + \frac{M^k}{N^2} \sum_{q=1}^{M-1} \sum_{j=0}^k
         \binom{k}{j} \frac{M^j}{q^{j-1}} \\
    &\le \mu + \frac{M^k}{N^2} \Bigg[ O(M^2 \log M) +  \sum_{j=3}^k
         \binom{k}{j} M^j \sum_{q=1}^\infty \frac{1}{q^{j-1}} \Bigg] \\
    &\le \mu + \frac{M^k}{N^2} \Bigg[ O(M^2 \log M) +
         \frac{\pi^2}{6} \sum_{j=3}^k \binom{k}{j} M^j \Bigg] \\
    &=   \mu + \frac{\pi^2}{6} + o(1)
\,.
\ea
where in the next to last step we have used the fact that
$\sum_{q=1}^\infty q^{-(j-1)} \le \pi^2/6$ for any $j \ge 3$.  As $\mu
= M^k/N =1 +o(1)$, we find $\sigma^2 = \E_{x,w}[(\eta^x_w)^2]-\mu^2
\le \pi^2/6 + o(1)$.

Since the variance is small, Chebyshev's inequality shows that the
probability of deviating far from the mean number of solutions is
small:
\be
  \Pr(|\eta^x_w - \mu| \ge \Delta) \le \frac{\sigma^2}{\Delta^2}
\,.
\label{eq:cheby}
\ee 
Putting $\Delta=4$ and using the fact that $\eta^x_w$ must be an
integer, we find $\Pr(\eta^x_w \ge 5) \le \pi^2/96 + o(1)$.

To see that we are unlikely to have no solutions, we need a slightly
stronger bound than the Chebyshev inequality.  Since $\eta^x_w\in\N$,
we have $\Pr(\eta^x_w = 0) \le {\sigma^2}/({\mu^2 + \sigma^2})$
\cite{AS00}*{p.~58}.  Now, noting that the $\gcd$ in
(\ref{eq:varlastexact}) is at least $1$, we have
\be
  \E_{x,w}[(\eta^x_w)^2]
    \ge \mu + \frac{M^k(M^k-1)}{N^2} = 2+o(1)
\ee
so that $\sigma^2 \ge 1 + o(1)$.  Therefore, we find $\Pr(\eta^x_w =
0) \le \pi^2/12 + o(1)$.  Combining these results, we see that $\Pr(1
\le \eta^x_w \le 4)\ge 1-3\pi^2/32 + o(1) \geq 0.0747 +o(1)$, so that
the probability is lower bounded by a constant for sufficiently large
$N$.
\end{proof}

While the above bounds apply to arbitrary values of $N$, they are not
tight, and better bounds can be obtained using knowledge of the
factorization of $N$.  For example, if $N$ is prime, $\sigma^2 \sim
1$.  For $k=2$, the above argument is not sufficient except for
special values of $N$ (such as $N$ prime); indeed, if $N$ has an
unbounded number of distinct prime factors, then it appears that the
variance of $\eta^x_w$ might be unbounded.  However, for this case,
one can simply decrease $M$ and use $k=3$ copies, as mentioned
previously.


\end{document}